\documentclass[11pt,oneside,english]{article}
\usepackage[a4paper, total={6.2in, 9in}]{geometry}
\usepackage[utf8]{inputenc}
\pdfoutput=1

\usepackage{graphicx}
\usepackage{amssymb}
\usepackage{xcolor}
\usepackage[T1]{fontenc}
\usepackage{lmodern}

\usepackage{caption}

\usepackage{courier}  

\usepackage[bottom, multiple]{footmisc}  
\usepackage[normalem]{ulem}    
\usepackage{anyfontsize}

\PassOptionsToPackage{hyphens}{url}\usepackage[hidelinks, hyperfootnotes=false]{hyperref}
\usepackage{fancyhdr}                           
\fancypagestyle{firstpage} 
{
   \fancyhf{}
   \fancyfoot[C]{\vspace{-12mm}\thepage}

   \fancyfoot[L]{\vspace{-5mm}\noindent\rule{6.2in}{0.4pt}
     \scriptsize{
         \textcopyright \hspace{0.5mm} Barclays Bank PLC 2021 \\
       This work is licensed under a \href{https://creativecommons.org/licenses/by/4.0/}{Creative Commons Attribution 4.0 International License (CC
       BY).}  \\ Provided you adhere to the CC BY license, including as to attribution, you are
       free to copy and redistribute this work in any medium or format and remix, transform,
       and build upon the work for any purpose, even commercially.                  \\
       BARCLAYS is a registered trademark of Barclays Bank PLC, all rights are reserved. \\
     }
 }
}

\title{\vspace{-2cm}Simulation of Derivatives Post-Trade Services using an Authoritative Data
  Store and the ISDA Common Domain Model}

\author{%
  \begin{tabular}{c} {\fontsize{10.75}{1cm}\selectfont Vikram A. Bakshi, Aishwarya Nair, Lee Braine} \\ {\fontsize{10.75}{1cm}\selectfont Chief Technology Office} \\
    {\fontsize{10.75}{1cm}\selectfont Barclays} \\ \hskip 1em \end{tabular} }

\date{\vspace{-0.5cm}October 6, 2021}

\begin{document}
\maketitle
\thispagestyle{firstpage} 
\vspace{-0.75cm}
\begin{abstract}
\noindent
In this paper, we present a summary of the design and implementation of a simulation of
post-trade services for interest rate swaps, from execution to maturity. We use an
authoritative data store (ADS) and the International Swaps and Derivatives Association (ISDA)
Common Domain Model (CDM) to simulate a potential future architecture.  We start by providing a
brief overview of the CDM and the lifecycle of an interest rate swap. We then compare our
simulated future state architecture with a typical current state architecture. Next, we present
the key requirements of the simulated system, several suitable design patterns, and a summary
of the implementation.  The simulation uses the CDM to address the industry problems of
inconsistent processes and inconsistent data, and an authoritative data store to address the
industry problem of duplicated data.
\end{abstract}

\section{Introduction}
\label{sec:introduction}

\noindent
The post-trade industry has gradually evolved over time, with its core infrastructure
increasing in complexity as new functionality is added. There is an opportunity to rationalise
this infrastructure and potentially reduce costs by 50 to 80 percent by adopting newer
standards and technologies \cite{deloitte-future-of-post-trade}.  In a previous paper
\cite{barclays-industry-adoption-scenarios-2020}, we highlighted that there exists ``a perfect
storm of industry inefficiency in post-trade processing, fuelled by duplicated inconsistent
processes operating on duplicated trade data in inconsistent data formats''. We also identified
a potential industry solution in the widespread adoption of the ISDA Common Domain Model (CDM)
\cite{isda-cdm-factsheet}, alongside the adoption of authoritative data stores (ADSs).  The
former provides common standards for both data and processes, and the latter addresses the
problem of data inconsistency and duplication. 

Here we aim to provide a summary report of relevance to financial institutions.  We set the
scene by presenting the scope, aim and structure of the CDM, as well summarising the lifecycle
of an interest rate swap. Our paper then provides an overview of the design and implementation
of a simulation of post-trade processing for interest rate swaps, from execution to
maturity. We compare our simulated future state architecture for trade services with a typical
current state architecture.  We also present the key requirements of the system, several
suitable architectural and software design patterns, and key implementation details.  Finally,
we briefly analyse the simulated architecture, including how it addresses the three industry
problems of inconsistent processes, inconsistent data, and duplicated data. 

This trade services simulation builds upon Barclays' existing engagement with the CDM,
including participating in association working groups, hosting industry hackathons
\cite{deloitte-future-of-post-trade, ey-barclays-derivhack-2019}, and producing an architecture
report \cite{barclays-industry-adoption-scenarios-2020}. We hope this paper will contribute to
the design landscape for post-trade infrastructure and look forward to continuing industry
collaboration on suitable architectures and the CDM.


\section{The ISDA Common Domain Model}
\label{sec:isda-cdm}

In this section, we provide an overview of the structure of the CDM with the intention of
providing relevant background knowledge for the subsequent implementation.

\subsection{Scope and Aim of the CDM}

ISDA, the International Securities Lending Association (ISLA), and the International Capital
Market Association (ICMA) signed a memorandum of understanding \cite{isda-icma-isla-mou-2021}
in August 2021 ``to strengthen collaboration on the future development of the CDM'' with the
aim of establishing a ``single, common digital representation of trade events and actions
across the lifecycle of financial products''.

The scope of the CDM has expanded from its initial focus on OTC derivatives in 2018
\cite{isda-publishes-cdm-2018} to a much larger remit including ``OTC derivatives, cash
securities, securities financing, and commodities'' \cite{cdm-rosetta-docs-overview}. The CDM
remains an open collaborative standard with the expectation that other standards setters and
associations will join with ISDA, ISLA and ICMA in future.

\subsection{Structure of the CDM}
\label{subsec:cdm-structure}
 
The CDM is defined in a domain specific language (DSL) called Rosetta
\cite{cdm-rosetta-overview}.\footnote{The Legend platform, hosted by the Fintech Open Source
Foundation (FINOS), can also be used for CDM development \cite{finos-cdm-case-study}.} Code
generators are available to convert source code (written in the Rosetta DSL) to other computer
languages (such as Java, Go, and DAML) so that users can adopt the CDM and build applications
on top of it in the language of their choice.\footnote{ The Java distribution of the CDM is
complete. Other distributions capture the data types and data structures of the whole model
but, in some cases, do not contain all the functional
specifications. \cite{cdm-rosetta-docs-overview}.}  CDM data structures largely mirror their
corresponding types in FpML but, in some cases, the model has been amended following the advice
of ISDA working groups.

The goal of a \textit{single} domain model is an ambitious objective, so it is important that
the CDM's structure be flexible enough to accommodate the diversity of the financial products
in scope.  In the following subsections, we briefly summarise the seven modelling dimensions of
the CDM, comprising three main dimensions (Product, Event, Process) and four other dimensions
(Legal Agreement, Reference Data, Namespace, Mapping) \cite{cdm-rosetta-docs-overview,
  cdm-rosetta-docs-modelling-dimensions}.

\subsubsection{Product Model}
 
In the CDM product model, the \texttt{TradableProduct} type represents an executable financial
product. It contains all of the associated data required for a trade including
\texttt{Counterparty} information, details of the \texttt{Product} itself, the
\texttt{PriceQuantity}, etc.

In the CDM, the qualification of a \texttt{Product} occurs by inferring its type from its
constituent components.  For example, a \texttt{Product} containing two
\texttt{InterestRatePayout} objects could be inferred to be a vanilla swap, whereas if it
contained an \texttt{InterestRatePayout} and an \texttt{EquityPayout} then it could be inferred
to be an equity swap instead.

The composition of CDM components permits a high level of component reusability and
qualification by inference provides flexibility (because the product model does not necessarily
need to be coupled to a specific naming convention or taxonomy).

\subsubsection{Event Model}

The CDM event model provides data structures to represent all possible lifecycle events
for a trade. It is organised into a four level hierarchy:

\begin{itemize}
  
  \item \textbf{Workflow.} Represents a set of actions that are required to trigger a business
    event. Applications can trigger business events directly or use workflows.

  \item \textbf{Business Event.} Represents an atomic lifecycle event of a trade, comprising
    one or more primitive events which must either all occur or none occur. For example, a
    partial novation business event must have both a \texttt{ContractFormationPrimitive} and a
    \texttt{QuantityChangePrimitive} occur.  The qualification of the type of a business event
    is inferred from the primitive events it contains, similar to the process for
    products.\footnote{In some cases, it may also be necessary to specify a relevant ``intent''
    qualifier to aid qualification.}
     
 \item \textbf{Primitive Event.} Defines an atomic change of state for a trade and is a
   building block component intended to form part of a business event.  Each primitive event
   contains a ``before'' and ``after'' \texttt{TradeState} permitting tracking of the lineage
   of the trade's state.

 \item \textbf{\texttt{TradeState}.} Represents the state of a trade throughout its entire
   lifecycle. The \texttt{TradeState} itself contains a \texttt{TradableProduct}, linking the
   event model to the product model.
\end{itemize}

\subsubsection{Process Model}

The CDM process model provides standardised representations of industry processes such as trade
execution, confirmation and settlement.\footnote{For the full list of processes in scope, see
the ``coverage'' section of the CDM documentation
\cite{cdm-rosetta-docs-process-model-coverage}.}  For each process, the CDM leverages the
function component of the Rosetta DSL to provide executable code for: (i) validation, (ii)
calculation (where applicable), and (iii) event creation.

Each function takes a CDM object as input and returns a CDM object as output.\footnote{Some
functions may also have additional basic types as input/output such as numbers, boolean values,
etc.}  This results in a standardised application programming interface (API) for automating
trade processing and increases the interoperability of components built using the CDM.


\subsubsection{Legal Agreement}

The CDM provides a digital representation of all types of ISDA documents (including Master
Agreement and Credit Support documentation) and business events are defined so that they can
reference the legal agreement(s) that govern them. Although not used in our implementation, it
should be noted that, by having digital representations of the ISDA legal agreements, there is
opportunity to increase efficiency in future legal and collateral workflows that adopt the CDM.

\subsubsection{Reference Data}

The reference data component contains definitions for parties and legal entities.

\pagebreak
\subsubsection{Namespace}

The Rosetta DSL files are organised into a hierarchy of namespaces intended to provide greater
modularity to the CDM.\footnote{The namespace hierarchy can be viewed graphically by browsing
the CDM using the Rosetta portal \cite{REGnosys-rosetta-portal}.} Components can reference each
other from anywhere in the model and the partitioning of components into namespaces has the
added benefits of shielding them from changes to each other and aiding understandability.

\subsubsection{Mapping (Synonym)}

The CDM defines synonym mappings between itself and existing standards, such as FpML, FIX, and
ISO 20022. These mappings aid in supporting both legacy applications and newer ``CDM native''
applications, as well as increasing interoperability.

\section{Lifecycle of an Interest Rate Swap}
\label{sec:irs-lifecycle}

In this section, we provide a high level overview of the lifecycle of a fixed versus floating
interest rate swap from execution to maturity,\footnote{We ignore pre-trade activities such as
counterparty onboarding and compliance checks as well as issues related to margin and
collateral.} as an introduction to the use case selected for our simulation.  We chose an
interest rate swap over other types of derivatives, because it is a simple, well understood
product that comprises the majority of the over-the-counter (OTC) derivative
market.\footnote{In 2019, interest rate swaps were reported to account for approximately sixty
percent of the total gross notional volume of OTC derivatives \cite{fontana2019anatomy}.}

A vanilla fixed versus floating interest rate swap is an agreement to exchange future cash
flows based on a fixed rate of interest with future cash flows based on a floating rate of
interest \cite{durbin2011}, with the main purpose being to support risk management by
institutions such as banks, brokers, dealers and corporations \cite{sacks2015}. The condensed
lifecycle summary which follows is drawn from the descriptions found in
\cite{boe-future-post-trade}, \cite{darbyshire2016}, \cite{baker2015}, \cite{shaik2014}, and
\cite{bis2013}.

\subsection{Execution}

The execution of a swap occurs when the trade's economic terms (and, optionally, specific
non-economic reference data) are agreed between both parties. Executions largely occur through
telephone calls (voice) or through an electronic communication network (ECN).\footnote{An
execution could potentially occur using other methods such as post, email, or in person
meetings but these occur less frequently.}

The typical economic terms that are required to be captured include: (i) the notional amount,
(ii) the trade execution and termination dates, (iii) the fixed interest rate, and (iv) the
floating rate index plus its tenor. In the case of a vanilla interest rate swap, many of the
additional economic terms are inferable using market conventions.\footnote{Specific economic
terms (such as the day count conventions, holiday calendars, and the spot lag) can be inferred
by market conventions \cite{opengamma2013}. }

The additional step of allocation, where the trade is assigned to the legal entity (or
sub-account of the legal entity) which will actually own it, takes place either with the
execution or afterwards during the post-trade process.

\pagebreak
\subsection{Post-Trade} 

\subsubsection{Trade Capture}

A ``trade ticket'' is created in the relevant post-trade processing system following the
successful execution of a trade. If the execution took place on an ECN then the ticket details
could potentially be populated automatically, whereas a voice executed trade would usually
require some form of manual input. Following the initial trade capture, the trade will also be
validated against a known set of rules and enriched with additional details (such as settlement
instructions) which are necessary for further processing.

\subsubsection{Matching and Confirmation}

For a bilaterally executed trade, once the trade details have been captured, both parties must
then attempt to match their respective views of the trade against each other on an electronic
matching platform. After matching takes place, they must then confirm that the recorded terms
are accurate between them. In contrast, when a trade has been executed on an ECN platform, the
platform will use its straight-through processing systems to handle the matching and
confirmation automatically.

The confirmation is the agreement of both parties to the terms of the trade.  The interest rate
swap will have an associated schedule of dates on which floating rate resets and payments
(discussed below) are triggered and its lifecycle continues until termination.

\subsubsection{Floating Rate Reset}
\label{subsec:floating-rate-reset}

The payment amount for each cash flow is calculated using the economic terms for the relevant
``leg'' of the trade, with one of the key details being that particular period's associated
interest rate. For the fixed leg, this rate is established when the trade is executed and the
value remains constant. For the floating leg, the rate is reset to the value which is observed
on the dates specified in the trade's schedule. Once reset, the rate remains fixed for the next
period.\footnote{The terms ``reset'' and ``fixing'' are sometimes used interchangeably to refer
to this process.}

\subsubsection{Clearing and Settlement}
\label{subsec:clearing-and-settlement}

The trade's schedule dictates the dates on which payment obligations are due and must be
settled between the parties in the swap.  ``Clearing'' denotes the process of performing all of
the required activities prior to the settlement and ``settlement'' refers to the actual
discharge of the contractual obligation (the payment) which, once performed, is irrevocable.

Depending on the agreement between the counterparties, the clearing process could potentially
involve, among other things, the netting, transmission, reconciliation and in some cases the
confirming of payment order instructions prior to the establishment of the final positions for
settlement.

Clearing is either ``bilateral'' or ``central'' with the former referring to the process being
performed directly between the parties to the trade and the latter referring to the process in
which the original trade is replaced by two new trades which contain the same economic terms
but with the parties now facing off to the central counterparty instead.\footnote{In practice,
the term ``clearing'', unless explicitly qualified, typically refers to central clearing.}

\subsection{Maturity} 
\label{subsec:maturity}

In the case where one of the parties to the trade would like to exit their position prior to
the maturity date (an ``early termination''), they could potentially elect to perform an
amendment, cancellation, offset or unwind of the trade.  Otherwise, at the point of the final
payment of the swap, all contractual obligations are discharged and the interest rate swap is
matured/terminated.


\section{Design}
\label{sec:design}

In this section, we start by comparing a typical current post-trade services architecture with the
proposed future state architecture that we simulated. We then discuss the key requirements for
our simulation and the design patterns that were adopted.

\subsection{Comparison of Post-Trade Services Architectures}

\begin{figure}[ht!]
  \begin{center}
  \includegraphics[width=1\linewidth]{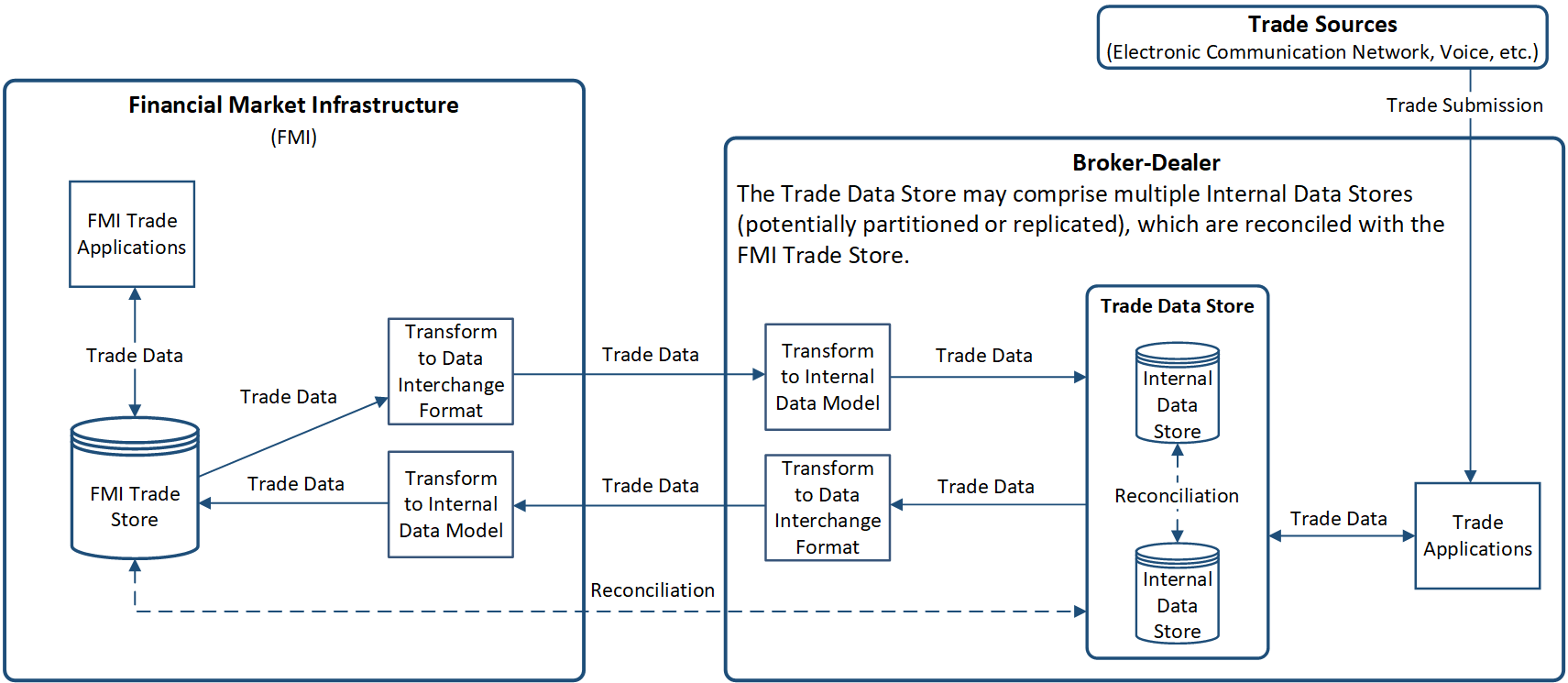}
  \end{center}
  \vspace{-4mm}
  \caption{\footnotesize{Current state - a typical post-trade services architecture.}}
  \label{fig:current-state-architecture}

\end{figure}

\begin{figure}[ht!]
  \begin{center}
  \includegraphics[width=1\linewidth]{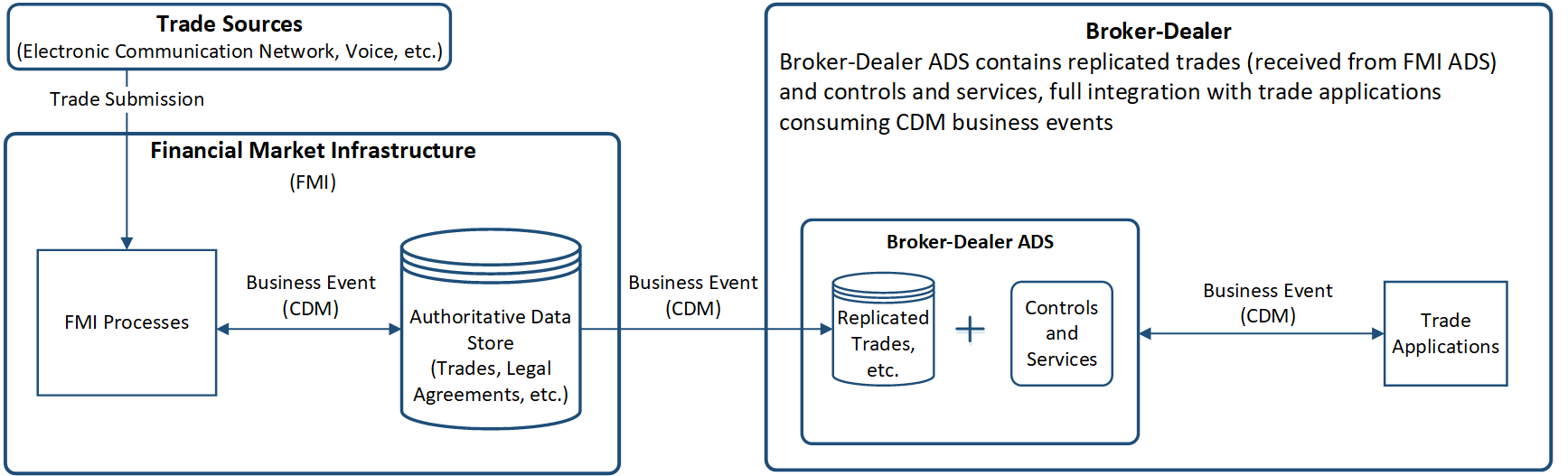}
  \end{center}
  \vspace{-4mm}
  \caption{\footnotesize{Future state - a central utility post-trade services architecture using
      the CDM.}}
  \label{fig:central-utility-ads-architecture}

\end{figure}

\pagebreak
\noindent
Figure \ref{fig:current-state-architecture} illustrates a typical current post-trade services
architecture in which a broker-dealer interacts with a financial market infrastructure (FMI).
Both parties utilise their own data models internally and therefore they must convert messages
to a data interchange format prior to sending them and also convert to their own internal data
format subsequent to receiving them.

The broker-dealer's applications can communicate with multiple trade stores containing data
that may be partitioned or replicated, and therefore may require reconciliation.  Additionally,
reconciliation is also typically required between between the trade stores of the FMI and the
broker-dealer to ensure consistency.

In contrast, Figure \ref{fig:central-utility-ads-architecture}\footnote{This architecture was
introduced in our previous paper as ``scenario 4''
\cite{barclays-industry-adoption-scenarios-2020}.}  illustrates the central utility
architecture in which both the FMI and broker-dealer utilise the CDM and, as a result, there is
no requirement for translations between data formats. The FMI hosts an ADS that contains all of
the associated lifecycle event data for trades. Broker-dealers can replicate the relevant
authoritative data from the ADS into their own local store, which can then be read by local
applications.

\subsection{Requirements}
\label{subsec:requirements}

Our simulation adopted the architecture illustrated in Figure
\ref{fig:central-utility-ads-architecture} and simulates the lifecycle of an interest rate swap
from execution to maturity. It comprises a central utility (hosting an ADS) and broker-dealers,
with all components using the CDM natively.  Given that the processes in post-trade services
will be simulated, we refer to our complete system as a ``trade services simulator''. The
following components are required:

\begin{itemize}
\item A centralised utility which acts as the FMI of the network, providing services for trade
  submission (execution) and consent (confirmation), as well as performing the actions required
  during the trade lifecycle (rate fixing and payments). Additionally, the FMI hosts the ADS
  containing the definitive record of the history of all trades.
\item A network operator which manages the processes of onboarding and offboarding participants
  on the network, and an associated identity registry of counterparties. The network operator
  functionality can be performed by the FMI, or an alternative entity.
\item A simulator ``harness'' containing a virtual network clock, whose time can be adjusted,
  and an associated scheduler which can trigger business events. A interest rate swap can be
  active for many years and, as such, the simulator harness is a key component which allows the
  full trade lifecycle to be performed on demand and within a practical time period.
\item A user interface allowing users to issue commands and queries to components in the
  system. For example, confirming that trade details are accurate or querying the FMI for the
  latest details of a trade.

\end{itemize}

\noindent
Note that netting and collateral management are out of scope for this trade services simulator.

\pagebreak
\subsection{Design Patterns}
\label{design-patterns}

A simulation of the lifecycle of an interest rate swap within a network of market participants
is a complex system. To tame this complexity, we adopted several suitable architectural and
software design patterns, which are discussed in the subsections below.

\subsubsection{Domain driven design (DDD)}

DDD\footnote{DDD terms are \textit{italicised} in the following description.} is a software
development approach in which the structure of the software matches the business domain
\cite{evans2004domain}. Developing software using DDD requires continually evolving your
\textit{domain model} such that it always remains an accurate representation of the real-world
processes in your domain.

A \textit{ubiquitous language} which is unambiguous and well understood within the domain is
used by both business experts and developers. Additionally, significant actions which can occur
within the domain are modelled as \textit{domain events}.

There is an explicit distinction made between \textit{value objects}, which are defined by
their attributes, and \textit{entities}, which are defined by their identifier.\footnote{An
example of a value object is a date - it is uniquely identifiable by its attributes (year,
month, day) and two variables containing the same date can be considered to be equal. An
example of an entity could be a book in a library - two instances of the exact same book would
not be considered as equal because they will have separate inventory identifiers/asset tags. }
Value objects and entities can be bound together to form an \textit{aggregate}, and other
domain objects can only refer to them through the identifier of the entity which is designated
as the \textit{aggregate root}.\footnote{An example of an aggregate could be a trade within the
CDM - the trade itself could be made up of several value objects and entities (e.g. payouts,
parties, etc), and all other objects can only refer to the trade through its unique identifier
(the root).}  The aggregate itself is the boundary for consistency, and changes in state are
applied such that invariants of the aggregate are always maintained. Retrieval of aggregates
occurs through the use of \textit{repositories} which fetch them from a store based on their
unique identifier.

In our case, our domain is that of post-trade processing and a significant amount of the
representation of the domain model is already provided by the CDM itself. As discussed in
section \ref{subsec:cdm-structure}, not only does the CDM provide representations for financial
products and business events, it goes further and provides representations of post-trade
processes as well. As a result, the CDM and DDD together make a complementary pairing for our
design and implementation.

\subsubsection{Event sourcing (ES)}

Event sourcing is a data storage pattern in which changes to an application's state are
captured as events which are committed to an append-only event store
\cite{event-sourcing-definition}.  The store is always the authoritative source of truth for
the system and provides a complete audit trail of all events. The events can also be
reprocessed at anytime to recreate the current state of the system.

The CDM already implements a form of event sourcing by tracking the effects of applying
business events in a ``before'' and ``after'' lineage contained within the event itself. This
provides the benefits of both traceability and visibility of state changes.  By adopting the
event sourcing pattern more widely, we can reap the associated benefits throughout the entirety
of the system and not just those parts which utilise the CDM.

\subsubsection{Command and query responsibility segregation (CQRS)}

CQRS is an architecture pattern which separates the handling of commands, which write data and
mutate state, and queries, which read data and do not mutate state
\cite{fowler-cqrs-definition}.

Events which are published by the command side are processed by the query side, which maintains
its own relevant ``projections'' of them into the desired view model.  This introduces
complexity because two models (command/query) must be maintained instead of one, and the
synchronisation of projections results in an eventually-consistent query side.

However, the benefits of using CQRS outweigh the drawbacks, especially given the complicated
domain of post-trade services. CQRS gives us the ability to: (i) develop components which are
optimised for a single purpose (command/query side), (ii) separately deploy and scale the
command and query sides based on their respective demand, (iii) select the most appropriate
storage solution for the command and query sides,\footnote{For example, in the CQRS
architecture we adopted for the FMI we used an event store for the command side and a
relational database for the query side.} and (iv) develop queries and projections for reports
or other data on an ad-hoc basis instead of requiring their specification in advance.

\subsubsection{Message driven architecture (MDA)}

A message driven architecture is a pattern in which components within a system communicate
with each other asynchronously through three types of message: commands, events, and queries. A
command message is a request for the system to perform some action (usually to mutate state),
an event message is a notification of an action having occurred (state having been mutated),
and a query message is a request for data to be returned to the sender.

Commands must be routed in a point-to-point manner from the component which is requesting the
state change directly to the DDD aggregate which it targets. The aggregate itself must
define a valid command handler for that particular command. The handler can then either fail
(for whatever reason) and raise an exception, or succeed and publish one or more events which
are then persisted on the event log and distributed to any event listeners.

In contrast to commands which are routed point-to-point, events follow the publish-subscribe
routing pattern where listeners receive published events without necessarily knowing anything
about the publisher itself.  This allows us to specialise parts of our system to follow an
event driven architecture with loose coupling as required.

\section{Implementation}
\label{sec:implementation}

In this section, we describe the technology stack, the architecture, and implementation details
for each of the four components of the system described earlier in section
\ref{subsec:requirements}.

\subsection{Technology Stack and Architecture}
 
The simulator is a loosely coupled system capable of managing the entire interest rate swap
lifecycle to maturity. It coordinates this through the mechanics of command processing and
event listeners, together with simulating moving forward through time.

We chose Java \cite{java-16} as the main programming language (together with the open-source
Spring Boot \cite{spring-boot} and Axon \cite{axon-ref-guide} frameworks) for development of
our backend services because of its complete integration with the CDM and our team's existing
familiarity with the language.  Spring Boot provides a rich environment of pluggable components
for testing, database connectivity and creating web endpoints, while the Axon framework
provides building blocks for the DDD, ES, CQRS, and MDA design patterns previously described in
section \ref{design-patterns}.

For the simulation, the backend data stack includes PostgreSQL \cite{postgresql}, an
open-source relational database, and Axon Server\footnote{Our simulation uses the open-source
standard edition of Axon Server.}  \cite{axon-ref-guide}, which acts as an event store. Axon
Server also acts as a message router for the system, simplifying the architecture by removing
the need for additional message-broker/queue infrastructure.

The frontend user interface uses the Vue.js \cite{vue-js} JavaScript framework, which was
selected over alternatives (such as Angular and React) because of its relative simplicity.

\begin{figure}[ht!]
\centerline{
   \includegraphics[width=1.15\linewidth]{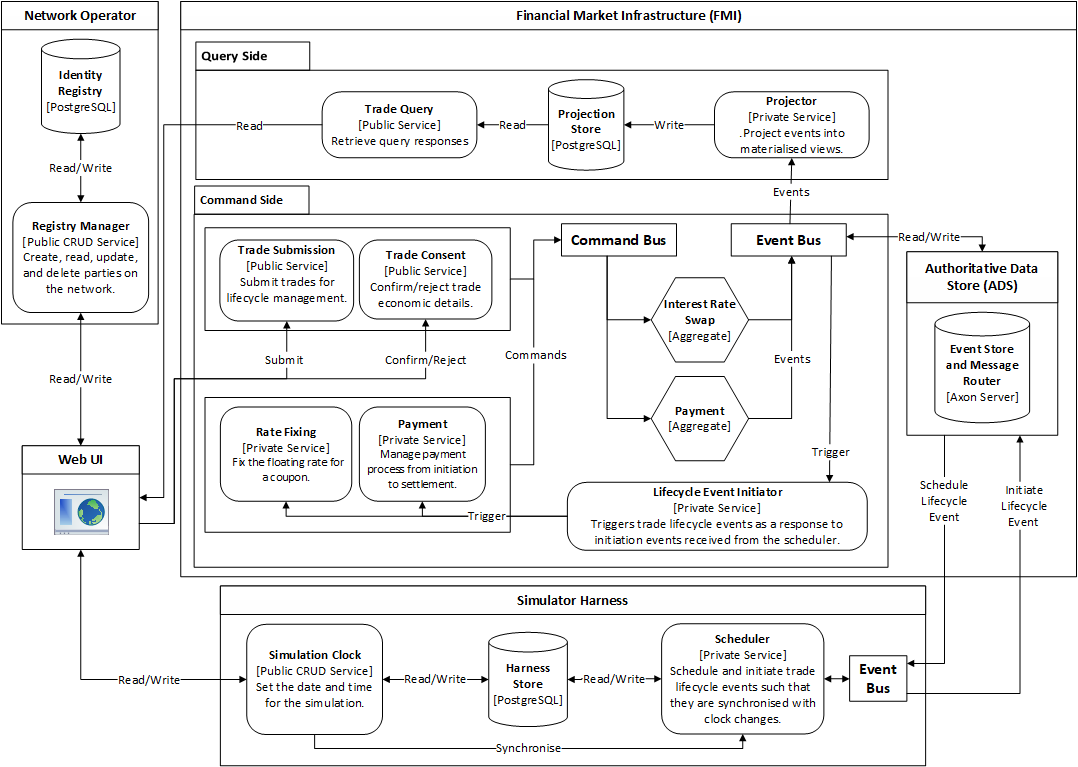}
}
  \caption{\footnotesize{Component architecture of the post-trade services simulator.}}
  \label{fig:simulator-architecture-diagram}

\end{figure}

Figure \ref{fig:simulator-architecture-diagram} illustrates the component architecture of the
post-simulator and the implementations of each of the constituent components are summarised below.


\vspace{1cm}


\pagebreak
\subsection{Network Operator}
\label{sec:network-operator}

In the network operator, a ``registry manager'' service exposes public interfaces to
create, read, update, and delete (CRUD) counterparties on the network, and a traditional
relational database (PostgreSQL) stores the counterparty data.

Users can utilise the Web UI to perform the required CRUD operations to ensure that their
required counterparties are present on the network.  In a real-world network, it is likely that
the FMI would act as the network operator, including hosting and managing the relevant
services.

\subsection{Web Based User Interface (UI)}
\label{sec:web-user-interface}

The web based UI contains four webpages which can be used to query the current state of the
system and to issue commands: simulator harness, network, interest rate swap execution, and
blotter. The UI allows the user to, for example, take the role of a broker-dealer via the
execution and blotter pages.

The simulator harness webpage provides functionality to start a new simulation, erasing any
pre-existing data. It can also be used to create a new clock for the current simulation and
initialise its date and time. The network webpage allows the user to interact with the registry
manager CRUD service to create, read, update, or delete parties from the network. The interest
rate swap execution webpage allows the user to submit a new trade execution to the FMI.

The fourth and final webpage, the blotter, communicates with the FMI's public query service and
displays key trade data including economic terms, open actions, and cashflow details.  If an
open action exists, for example the need to confirm an execution, then the webpage gives the
user the option to respond and close the action.

\begin{figure}[ht!]
\centerline{
   \includegraphics[width=1.1\linewidth]{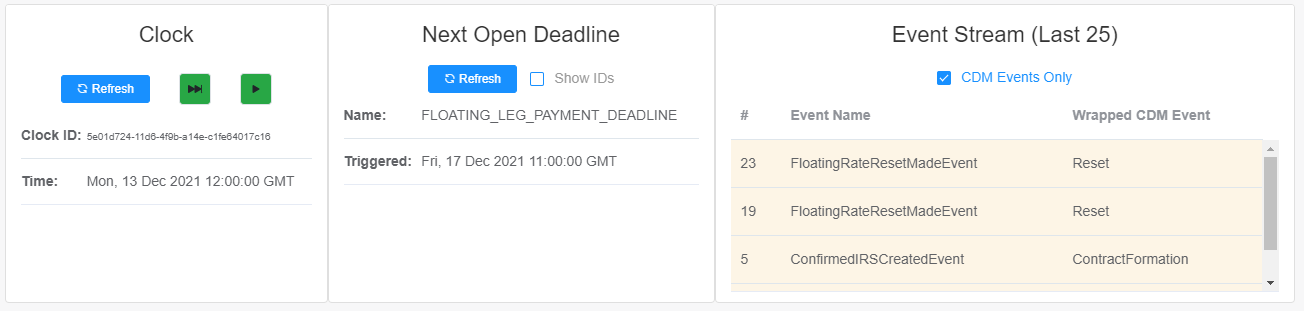}
}
  \caption{\footnotesize{The three widgets in the header of the blotter webpage.}}
  \label{fig:blotter-header-widgets}

\end{figure}

\vspace{-2mm}


\noindent
The header on the blotter webpage displays three widgets, as shown in Figure
\ref{fig:blotter-header-widgets}. The first widget, the clock, interacts with the simulation
clock backend service to: display the current simulator time, provide a ``forward'' button
which moves time forward to trigger the next open deadline, and provide a ``play'' button which
continues moving time forward until no open deadlines remain.

A ``deadline'' is the term used within the simulator to refer to a point in time when a trade
business event should be triggered. The second widget displays the name and time that the next
open deadline should be triggered. In this example, on the 17th December, there is a floating
leg payment which needs to be triggered, and clicking the forward button on the clock widget
would move time forward to this date.

The final widget is the event stream which shows the last twenty-five events which have
occurred in the current simulation run. In this particular case the filter is selected so that
only CDM events are displayed and other simulation events, such as the clock moving forward in
time, are not shown. The first column is the number of the event in the stream, the second
column is the name of the simulator event, and the last column is the underlying CDM event
which it represents. The underlying CDM event is wrapped into a simulator event so that it is
in a form which can be routed to other event listener components within the system. A
description of each of the different types of CDM events is provided in Table
\ref{table:irs-cdm-business-events}.

\subsection{Simulator Harness}

The harness component of the simulator comprises two services which persist their data to a
PostgreSQL relational database. The first service, the simulation clock, exposes public CRUD
endpoints which are then utilised by the UI to set the time and date of the simulation clock.

The second service, the scheduler, registers itself as a listener for all events which require
an action to be scheduled in the future and also as a listener for any clock change events so
that the scheduler can remain synchronised with the simulation clock. At the point when the
clock's time advances on or beyond the same time as a scheduled action, it publishes an event
to indicate that the scheduling threshold has been breached and the action should take
place. The lifecycle event initiator service in the FMI listens for these types of events and
reacts by triggering the relevant trade lifecycle event.

\subsection{Financial Market Infrastructure}

The FMI provides the key trade services within the simulator.  Its use of the CQRS pattern can
be seen in Figure \ref{fig:simulator-architecture-diagram} with each of the FMI's subcomponents
segregated into either the command side or the query side. A newly executed trade can be
submitted to the FMI's submission service using the interest rate swap execution webpage in the
UI. The user can then view the trade on the blotter webpage and issue a command (that is routed
to the FMI's consent service) to confirm or reject the accuracy of the trade details.  These
two services, submission and consent, manage the trade's execution and confirmation lifecycle
steps respectively.

Each of the commands that are issued are placed on the command bus which routes them to the
correct instance of the interest rate swap aggregate based on the trade identifier. The
aggregate contains the core business logic that validates that the command is executable, and
publishes event(s) indicating that the state has been successfully changed.

A number of events are published when the trade is confirmed, including those indicating the
dates on which future rate fixings and payments should be scheduled. Published events are
placed onto the event bus which persists them in the event store and then routes them to any
simulator component which is listening for those types of events.

The query side projector service listens for relevant events and reacts to them by maintaining
its materialised views\footnote{A materialised view is a logical data view which is persisted
to a database table.} in the PostgreSQL projection store. Additionally, the simulator harness
scheduler listens for any specific scheduling events which it will need to manage.  The
scheduler remains synchronised with the clock and each time the clock crosses a time threshold
it publishes an event to trigger the occurrence of the relevant business event by the FMI.

The FMI's lifecycle event initiator service listens for these types of events and triggers
either a rate fixing or payment to be made as a response. The rate returned from a rate fixing
is currently set to be a randomly generated number, and settlement of a payment is managed by a
payment aggregate which currently only simulates the action. However, the interfaces for these
actions are clearly defined and the current implementations could potentially be swapped in
future for ones which perform the genuine action.

\subsubsection{CDM usage}

The FMI manages the interest rate swap aggregate and makes extensive use of the CDM. The CDM
product model, with the root component being the \texttt{TradableProduct} type, is used to
define an interest rate swap and its \texttt{EconomicTerms}.

Table \ref{table:irs-cdm-business-events} describes the four business events (from the CDM
event model) which occur during the lifecycle of the interest rate swap and are used by the
FMI.

\begin{table}[ht!]
  \centerline{
    \begin{tabular}{| l | p{10.5cm} |}
      \hline
      
      \multicolumn{1}{|c|}{\textbf{CDM Business Event}} &
      \multicolumn{1}{|c|}{\textbf{Description}} \\ \hline

      Execution         & Contains a trade whose \texttt{EconomicTerms} are waiting to be confirmed by the two parties.         \\ \hline
      ContractFormation & Contains an executed trade which has been confirmed by the two parties.                             \\ \hline
      Reset             & Contains a trade whose last event was a reset of the associated rate within the floating rate payout. \\ \hline
      CashTransfer      & Contains a trade whose last event was a transfer of a cash amount between the two parties.          \\ \hline
    \end{tabular}
  }
  \caption{CDM business events for an interest rate swap.}
  \label{table:irs-cdm-business-events}
  \end{table}


\noindent
Integrating the CDM process model is not straightforward because it is necessary to determine
whether the default function implementation(s) in the CDM distribution will suffice for the
given use case or whether further customisation is required.

For our simulator, the FMI contains a customised implementation of a business event creation
function for each business event described in Table \ref{table:irs-cdm-business-events}.  These
in turn make use of the default CDM implementations for generating the event primitives that
are contained within the final business event. When the specific point in the trade's lifecycle
is reached, the correct function is invoked and a validated and qualified business event of
that type is created.

Additionally, the FMI contains a custom implementation of the \texttt{ResolveObservation}
function which is invoked when an observation of the underlying reference rate for the floating
leg needs to be made. 

\section{Architecture Analysis}

The three fundamental problems facing the post-trade industry were summarised in
\cite{barclays-industry-adoption-scenarios-2020} as being inconsistent processes, inconsistent
data, and duplicated data. Developing a working implementation of the system illustrated in
Figure \ref{fig:simulator-architecture-diagram} has provided insights on how these industry
problems could be addressed by using an authoritative data store with the CDM.

In our simulation, processes cannot be inconsistent because they are defined by the CDM and
implemented by the FMI, which drives adoption by the broker-dealers, thereby removing any
potential for variation. Similarly, data cannot be inconsistent because the CDM provides common
definitive representations of business events and data, and all market participants use the
same authoritative data store. This authoritative data store also removes inefficiencies, such
as reconciliation, resulting from trade data being duplicated across multiple locations.

The simulated future state architecture is also less complex than a typical current state
architecture, as can be seen by comparing Figures \ref{fig:current-state-architecture} and
\ref{fig:central-utility-ads-architecture}. The encapsulation of almost all of the trade
processes within a single entity (the FMI utility) is more efficient than each of the
broker-dealers performing those processes themselves. This provides an opportunity for
rationalisation of hardware/software infrastructure and cost reduction.

By utilising the CDM throughout the system, we avoid issues associated with standards
proliferation, such as the need to maintain mappings, the challenges with interoperability, and
the potential risk of data loss. In contrast, note that a typical trade flow could currently be
``executed using the pre-trade FIX protocol (with an FpML payload representing the product),
confirmed electronically using FpML as the contract representation, and reported to a Trade
Repository under the ISO 20022 format'' \cite{cdm-rosetta-docs-overview}.

\section{Summary and Further Work}
\label{subsec:summary-and-further-work}

This paper provided an overview of the design and implementation of a simulation of post-trade
services for interest rate swaps, from execution to maturity. We began by presenting the scope,
aim and structure of the CDM. We then summarised the lifecycle of an interest rate swap
including the stages of execution, matching, confirmation, floating rate reset, clearing,
settlement and maturity. 

We then moved on to compare our simulated future state architecture for trade services with a
typical current state architecture.  Next, we presented the key requirements of the system and
several suitable design patterns which we adopted.  We then summarised the implementation
including the technology stack, the architecture, and details of for each of the components of
the system.

Finally, we briefly analysed the simulated architecture by considering how it addressed the
three fundamental problems facing the post-trade industry. This included using the CDM to
address the industry problems of inconsistent processes and inconsistent data, and using
authoritative data stores to address the industry problem of duplicated data. We also
highlighted the opportunity for this less complex architecture to rationalise infrastructure
and reduce costs in the post-trade industry.

We have demonstrated that the CDM is capable of modelling the entire lifecycle of a trade from
execution to maturity and can report that our exploration of the central utility architecture
has validated the opportunity for simplification of the post-trade industry.

Potential further work includes incorporating legal agreements into the simulation's workflows,
storing supplementary private data for a trade in the ADS, and providing digital regulatory
reporting. We look forward to continuing industry collaboration on the CDM and opportunities to
improve the post-trade industry.

\vspace{5mm}

\noindent \textbf{ACKNOWLEDGEMENTS:} 
\noindent
We would like to thank Jim Wang (REGnosys), Ruddy Vincent (ISDA) and Ted Sandborn (ISDA) for
implementation support. Thanks are also due to Rajagopalan Siddharthan (Barclays), Mark
Gladstone (Barclays) and Ian Sloyan (ISDA) for their helpful feedback. 

\pretolerance=-1
\tolerance=-1
\emergencystretch=0pt

\bibliography{ads-cdm-simulator-paper}
\bibliographystyle{plain}

\end{document}